\documentstyle[12pt]{article}

\catcode`\@=11
\def\@citex[#1]#2{%
\if@filesw \immediate \write \@auxout {\string \citation {#2}}\fi
\@tempcntb\m@ne \let\@h@ld\relax \def\@citea{}%
\@cite{%
  \@for \@citeb:=#2\do {%
    \@ifundefined {b@\@citeb}%
      {\@h@ld\@citea\@tempcntb\m@ne{\bf ?}%
      \@warning {Citation `\@citeb ' on page \thepage \space undefined}}%
      {\@tempcnta\@tempcntb \advance\@tempcnta\@ne%
      \@tempcntb\number\csname b@\@citeb \endcsname \relax%
      \ifnum\@tempcnta=\@tempcntb 
	\ifx\@h@ld\relax%
	  \edef \@h@ld{\@citea\csname b@\@citeb\endcsname}%
	\else%
	  \edef\@h@ld{\ifmmode{-}\else--\fi\csname b@\@citeb\endcsname}%
	\fi%
      \else
	\@h@ld\@citea\csname b@\@citeb \endcsname%
	\let\@h@ld\relax%
      \fi}%
    \def\@citea{,\penalty\@highpenalty\,}%
  }\@h@ld
}{#1}}

\def\@citeb#1#2{{[#1]\if@tempswa , #2\fi}}
\def\@citeu#1#2{{$^{#1}$\if@tempswa , #2\fi }}
\def\@citep#1#2{{#1\if@tempswa , #2\fi}}

\def\bcites{         
	\catcode`\@=11
	\let\@cite=\@citeb
	\catcode`\@=12
}

\def\upcites{         
	\catcode`\@=11
	\let\@cite=\@citeu
	\catcode`\@=12
}

\def\plaincites{      
	\catcode`\@=11
	\let\@cite=\@citep
	\catcode`\@=12
}

\newcount\hour
\newcount\minute
\newtoks\amorpm
\hour=\time\divide\hour by 60
\minute=\time{\multiply\hour by 60 \global\advance\minute by-\hour}
\edef\standardtime{{\ifnum\hour<12 \global\amorpm={am}%
	\else\global\amorpm={pm}\advance\hour by-12 \fi
	\ifnum\hour=0 \hour=12 \fi
	\number\hour:\ifnum\minute<10 0\fi\number\minute\the\amorpm}}
\edef\militarytime{\number\hour:\ifnum\minute<10 0\fi\number\minute}

\def\draftlabel#1{{\@bsphack\if@filesw {\let\thepage\relax
   \xdef\@gtempa{\write\@auxout{\string
      \newlabel{#1}{{\@currentlabel}{\thepage}}}}}\@gtempa
   \if@nobreak \ifvmode\nobreak\fi\fi\fi\@esphack}
	\gdef\@eqnlabel{#1}}
\def\@eqnlabel{}
\def\@vacuum{}
\def\marginnote#1{}
\def\draftmarginnote#1{\marginpar{\raggedright\scriptsize\tt#1}}
\def\draft{
	\pagestyle{plain}
	\overfullrule=2pt
	\oddsidemargin -.5truein
	\def\@oddhead{\sl \phantom{\today\quad\militarytime} \hfil
	\smash{\Large\sl DRAFT} \hfil \today\quad\militarytime}
	\let\@evenhead\@oddhead
	\let\label=\draftlabel
	\let\marginnote=\draftmarginnote
	\def\ps@empty{\let\@mkboth\@gobbletwo
	\def\@oddfoot{\hfil \smash{\Large\sl DRAFT} \hfil}
	\let\@evenfoot\@oddhead}
	\def\@eqnnum{(\theequation)\rlap{\kern\marginparsep\tt\@eqnlabel}%
	\global\let\@eqnlabel\@vacuum}  }

\def\blackfonts{
	\font\blackboard=msbm10 scaled\magstep1
	\font\blackboards=msbm8
	\font\blackboardss=msbm6
}

\def\nblack{            
	\def\ZZ{{Z \n{10} Z}}
	\def\NN{{N \n{14} N}}
	\def\CC{{C \n{11} C}}
	\def\RR{{R \n{11} R}}
	\def\QQ{{Q \n{12} Q}}
	\def\PP{{P \n{11} P}}
}

\def\prep{         
	\catcode`\@=11
	\input art10.sty
	\catcode`\@=12
	
	\let\small\null
	\def\blackfonts{
		\font\blackboard=msbm10
		\font\blackboards=msbm7
		\font\blackboardss=msbm5
	}
	\let\sl\it
	\twocolumn
	\sloppy
	\voffset=-2.54truecm
	\hoffset=-2.54truecm
	\flushbottom
	\parindent 1em
	\leftmargini 2em
	\leftmarginv .5em
	\leftmarginvi .5em
	\marginparwidth 48pt
	\marginparsep 10pt
	\setlength{\columnsep}{2truecm}
	\setlength{\textwidth}{25.4truecm}
	\setlength{\textheight}{17truecm}
	\baselineskip=16pt
	\oddsidemargin .18truein
	\evensidemargin .17truein
}

\def\eqalign#1{\null\,\vcenter{\openup\jot\m@th
  \ialign{\strut\hfil$\displaystyle{##}$&$\displaystyle{{}##}$\hfil
      \crcr#1\crcr}}\,}
\def\eqalignno#1{\displ@y \tabskip\centering
  \halign to\displaywidth{\hfil$\@lign\displaystyle{##}$\tabskip\z@skip
    &$\@lign\displaystyle{{}##}$\hfil\tabskip\centering
    &\llap{$\@lign##$}\tabskip\z@skip\crcr
    #1\crcr}}

\def\section{\@startsection {section}{1}{\z@}{3.ex plus 1ex minus
 .2ex}{2.ex plus .2ex}{\large\bf}}
\def\subsection{\@startsection{subsection}{2}{\z@}{2.75ex plus 1ex minus
 .2ex}{1.5ex plus .2ex}{\bf}}

\def\appendix{{\newpage\section*{Appendices}}\let\appendix\section%
	{\setcounter{section}{0}
	\gdef\thesection{\Alph{section}}}\section}

\def\abstract{\if@twocolumn
\section*{Abstract}
\else 
\begin{center}
{\bf Abstract\vspace{-.5em}\vspace{0pt}}
\end{center}
\quotation
\fi}

\catcode`\@=12

\def\noj#1,#2,{{\bf #1} (19#2)\ }
\def\jou#1,#2,#3,{{\sl #1\/ }{\bf #2} (19#3)\ }
\def\ann#1,#2,{{\sl Ann.\ Physics\/ }{\bf #1} (19#2)\ }
\def\cmp#1,#2,{{\sl Comm.\ Math.\ Phys.\/ }{\bf #1} (19#2)\ }
\def\cq#1,#2,{{\sl Class.\ Quantum Grav.\/ }{\bf #1} (19#2)\ }
\def\cqg#1,#2,{{\sl Class.\ Quantum Grav.\/ }{\bf #1} (19#2)\ }
\def\ijmp#1,#2,{{\sl Int.\ J.\ Mod.\ Phys.\/ }{\bf A#1} (19#2)\ }
\def\jmp#1,#2,{{\sl J.\ Math.\ Phys.\/ }{\bf #1} (19#2)\ }
\def\grg#1,#2,{{\sl Gen.\ Rel.\ Grav.\/ }{\bf #1} (19#2)\ }
\def\mpl#1,#2,{{\sl Mod.\ Phys.\ Lett.\/ }{\bf A#1} (19#2)\ }
\def\nc#1,#2,{{\sl Nuovo Cim.\/ }{\bf #1} (19#2)\ }
\def\np#1,#2,{{\sl Nucl.\ Phys.\/ }{\bf B#1} (19#2)\ }
\def\pl#1,#2,{{\sl Phys.\ Lett.\/ }{\bf #1B} (19#2)\ }
\def\pla#1,#2,{{\sl Phys.\ Lett.\/ }{\bf #1A} (19#2)\ }
\def\pr#1,#2,{{\sl Phys.\ Rev.\/ }{\bf #1} (19#2)\ }
\def\prd#1,#2,{{\sl Phys.\ Rev.\/ }{\bf D#1} (19#2)\ }
\def\prl#1,#2,{{\sl Phys.\ Rev.\ Lett.\/ }{\bf #1} (19#2)\ }
\def\prp#1,#2,{{\sl Phys.\ Rept.\/ }{\bf #1C} (19#2)\ }
\def\ptp#1,#2,{{\sl Prog.\ Theor.\ Phys.\/ }{\bf #1} (19#2)\ }
\def\ptpsup#1,#2,{{\sl Prog.\ Theor.\ Phys.\/ Suppl.\/ }{\bf #1} (19#2)\ }
\def\rmp#1,#2,{{\sl Rev.\ Mod.\ Phys.\/ }{\bf #1} (19#2)\ }
\def\yadfiz#1,#2,#3[#4,#5]{{\sl Yad.\ Fiz.\/ }{\bf #1} (19#2) #3%
\ [{\sl Sov.\ J.\ Nucl.\ Phys.\/ }{\bf #4} (19#2) #5]}
\def\zh#1,#2,#3[#4,#5]{{\sl Zh.\ Exp.\ Theor.\ Fiz.\/ }{\bf #1} (19#2) #3%
\ [{\sl Sov.\ Phys.\ JETP\/ }{\bf #4} (19#2) #5]}

\def\beq{\begin{equation}}
\def\eeq{\end{equation}}
\def\beqar{\begin{eqnarray}}
\def\eeqar{\end{eqnarray}}

\def\nfrac#1#2{{\displaystyle{\vphantom1\smash{\lower.5ex\hbox{\small$#1$}}%
	\over\vphantom1\smash{\raise.25ex\hbox{\small$#2$}}}}}

\def\p#1{\mskip#1mu}
\def\n#1{\mskip-#1mu}
\def\stop{\p6.}
\def\comma{\p6,}

\def\lae{\mathrel{\mathop{\smash{\lower .5 ex \hbox{$\stackrel<\sim$}}}}}
\def\lae{\mathrel{\mathop{\smash{\lower .5 ex \hbox{$\stackrel>\sim$}}}}}

\def\Tr{{\rm Tr}}
\def\l:{\mathopen{:}\,}
\def\r:{\,\mathclose{:}}

\def\[{\left[}          \def\]{\right]}
\def\({\left(}          \def\){\right)}
\def\<{\left<}          \def\>{\right>}

\def\CF{{\cal F}}
\def\CL{{\cal L}}

\def\CR{{\cal R}}

\def\mod{{\rm mod}}

\catcode`\@=11
\def\theequation{\arabic{equation}}
\catcode`\@=12

\nblack
\bcites

\nblack

\catcode`\@=11
\def\theequation{\thesection.\arabic{equation}}
\@addtoreset{equation}{section}
\@addtoreset{footnote}{section}
\@addtoreset{footnote}{subsection}
\catcode`\@=12

\newcommand{\beqn}{\begin{equation}}
\newcommand{\eeqn}{\end{equation}}
\newcommand{\beqnarray}{\begin{eqnarray}}
\newcommand{\eeqnarray}{\end{eqnarray}}

\newcommand{\res}{\;\mathop{\mbox{\rm res}}}
\begin{document}
\begin{titlepage}

\begin{center}
\today
\hfill       TAUP--2248--95 \\
\hfill                  WIS-95/19/May-PH               \\
\hfill                  hep-th/9505075

\vskip 1 cm
{\large \bf On the
Quantum Moduli Space of Vacua of $N=2$ Supersymmetric
$SU(N_c)$ Gauge Theories\\}
\vskip 0.1 cm
\vskip 0.5 cm
{Amihay Hanany}
\vskip 0.2cm
{\sl
ftami@wicc.weizmann.ac.il \\
Department of Particle Physics \\
Weizmann Institute of Science \\
 76100 Rehovot Israel
}
\vskip 0.2cm
{Yaron Oz\footnote{Work
supported in part by the US-Israel Binational Science Foundation,
and the Israel Academy of Science.}
}
\vskip 0.2cm

{\sl
yarono@ccsg.tau.ac.il \\
School of Physics and Astronomy\\Raymond and Beverly Sackler Faculty
of Exact Sciences\\Tel-Aviv University\\Ramat Aviv, Tel-Aviv 69978, Israel.
}

\end{center}

\vskip 0.5 cm
\begin{abstract}
We construct families of hyper-elliptic curves which describe the quantum
moduli
spaces of vacua of $N=2$ supersymmetric $SU(N_c)$ gauge theories
coupled to $N_f$ flavors of quarks in the fundamental representation.
The quantum moduli spaces for $N_f < N_c$ are
determined completely by imposing $R$-symmetry,
instanton corrections and
the proper classical singularity structure. These curves are verified
by residue and weak coupling monodromy calculations.
The quantum moduli spaces
for $N_f\geq N_c$ theories are parameterized and their general
structure is worked out using
residue calculations.
Global symmetry considerations suggest a complete description
of them. The results are supported by weak coupling monodromy
calculations.
The exact metrics on the quantum moduli spaces as well as the exact
spectrum of stable massive states are derived.
We find an example of a novel symmetry of a quantum moduli
space: Invariance under the exchange of a moduli parameter
and the bare mass.
We apply our method for the construction of the
quantum moduli spaces of vacua
of $N=1$ supersymmetric theories in the coulomb phase.
\end{abstract}

\end{titlepage}

\section{Introduction}

There has been much progress recently in the study of
non-perturbative properties of $N=1$
and $N=2$
four dimensional supersymmetric field theories.
Of particular importance to the analysis
is the concept of holomorphy \cite{Natiholo}.
In the $N=1$ theories, the super-potential and the coefficients of
the gauge kinetic terms are holomorphic and are therefore much
constrained.
In the $N=2$ theories, the K\"ahler potential is also constrained
by holomorphy, thus making the holomorphy even more powerful.

A basic ingredient in the analysis is the moduli space
of vacua corresponding to a continuous degeneracy of inequivalent
ground states.
Classically, the super-potential has flat directions along which the squarks
get vacuum expectation values and thus break the gauge symmetry.
The singularities of the classical moduli space are of two types:
They either correspond to
enhanced gauge symmetry where the gauge group is not broken to its
maximal torus, or
correspond to vacua where some of the matter fields become
massless.

The quantum moduli space is in general different from the classical moduli
space.
In \cite{sw1} the quantum moduli space of $N=2$ supersymmetric
pure $SU(2)$ gauge theory was determined as a certain
one parameter family
of elliptic curves. The exact metric on the quantum moduli space
as well as the exact spectrum of BPS particles were found
via the periods of the curves.
The two singularities of the quantum moduli space were associated
with massless monopole and dyon.
The generalization of \cite{sw1} to include massless and
massive matter hyper-multiplets
was carried out in \cite{sw2}.
The quantum moduli space of $N=2$ supersymmetric pure $SU(N_c)$
gauge theories was constructed as an $N_c-1$
parameter family of hyper-elliptic curves in \cite{Shimon,Alon},
and the physics of the model was studied in \cite{DS}.

Our aim in this paper is to determine the quantum moduli spaces of vacua of
the coulomb phase of
$N=2$ supersymmetric $SU(N_c)$ gauge theories with $N_f$ matter
hyper-multiplets in the fundamental representation, as well as
the exact metric on the quantum moduli space and the exact spectrum of
massive stable particles.
The quantum moduli spaces will be constructed as families of hyper-elliptic
curves satisfying a set of physical constraints: $R$-symmetry,
singularity structure and global symmetries,
appropriate inclusion of instanton correction, proper classical
and scaling (integration of massive quarks) limits,
compatibility of residue calculations with the BPS formula and
correct weak coupling monodromies.

The paper is organized as follows. In section two
we review general aspects of $N=2$ gauge
theories and the moduli spaces of vacua,
introduce the physical quantities
associated with them,
discuss their singularity
and monodromy structures,
and present principles for their construction.
In section three we construct the quantum moduli space for $N_f < N_c$.
We show that it is determined completely by imposing $R$-symmetry,
instanton corrections and
the proper classical singularity structure.
We get compatibility of the residue calculations with the BPS formula
which provides a consistency check of the result.
Section four is devoted to the $N_c=N_f$ theories.
Applying the physical constraints including residue calculations
leaves us with one undetermined constant parameter. In the $SU(2)$ case it is
determined by imposing an appropriate $Z_2$ symmetry in the moduli space.
For $N_c > 2$ we determine the parameter by requiring compatibility,
upon integration of a massive
quark,
with the $N_f> N_c$ theories which are worked out in the subsequent
section.

In section five the quantum moduli spaces
for $N_f > N_c$ theories are parameterized and their general
structure is worked out using
residue calculations. This still leaves us with undetermined
constant coefficients.
Global symmetry considerations suggest a complete determination
of them, which we conjecture but do not prove.

The results of the previous sections are supported by weak coupling
monodromy calculations in section six.
In section seven we discuss the $N_c=2N_f$ theories.
When the bare masses are zero one expects scale invariant theories
with periods satisfying the classical relations.
The quantum moduli spaces in the massless as well as
the massive theories are parameterized and their general
structure is worked out using
residue calculations in a similar manner to the
$N_f>N_c$ case.
Again, global symmetry considerations suggest a complete description
of the quantum moduli spaces which we do not prove.
The examples of quantum moduli spaces for $N_c=2,3$ are
worked out in detail in each section.

In general our results provide the exact metrics on the
quantum moduli spaces as well as the exact
spectrum of stable massive states.
Along the way we find an example of a novel symmetry of a quantum moduli
space: Invariance under the exchange of a moduli parameter
and the bare mass.
Section eight is devoted to discussion and conclusions.

The method that we use in order to construct the quantum moduli spaces
of $N=2$ theories is rather general. It can be used in order to
construct the quantum moduli spaces of a large class of $N=1$ theories,
in the coulomb phase, with different matter content.
In the appendix, as an example, we apply our method for constructing the
quantum moduli spaces, to $N=1$ supersymmetric $SU(N_c)$ gauge theories
with a single matter field in the adjoint representation
of the gauge group and $N_f$ matter fields in the fundamental
representation. The results generalize those of \cite{EFGR}
to $N_c>2$.

\section{The moduli space of vacua of $N=2$ gauge theories}
In this section we review general aspects of $N=2$ gauge
theories and the moduli spaces of vacua,
introduce the physical quantities
associated with them,
discuss their singularity
and monodromy structures,
and present principles for their construction.

\subsection{$N=2$ QCD and the moduli space of vacua}

We will consider $N=2$ supersymmetric $SU(N_c)$ gauge theories
with $N_c$ colors and $N_f$ flavors.
The field content of the theories consists of $N=2$ chiral multiplet
and $N_f$ hyper-multiplets.
The $N=2$ chiral multiplet contains gauge fields $A_{\mu}$, two
Weyl fermions $\lambda$ and $\psi$ (alternatively one Dirac fermion) and a
complex scalar $\phi$, all in the
adjoint representation of the gauge group $SU(N_c)$.
The $N=2$ hyper-multiplets that we will consider
contain two Weyl fermions $\psi_q$ and $\psi_{\tilde{q}}^{\dagger}$ and two
complex bosons $q$ and $\tilde{q}^{\dagger}$, all in the fundamental
representation of $SU(N_c)$.

In terms
of $N=1$ super-fields the $N=2$
chiral multiplet consists of a vector multiplet $W_{\alpha}$ and a
chiral multiplet  $\Phi$, while the $N=2$
hyper-multiplets consist of two chiral super-fields
$Q^i{}_a$ and $\tilde Q_i{}_a$, where $i=1,\dots,N_f$ is
the flavor index and
$a=1,\dots,N_c$ is the color index.
The super-potential in the $N=1$ language reads
\beq
W=\sqrt {2} \tilde Q_i \Phi Q^i + \sum_i m_i\tilde Q_i Q^i  \comma
\label{superpot}
\eeq
with $m_i$ being the bare masses and color indices are suppressed.
The first term in (\ref{superpot}) is related  by $N=2$
super-symmetry to the gauge couplings and the second term corresponds
to $N=2$ invariant mass terms.
When the quarks are massless the global symmetry of the classical theory is a
certain subgroup of
$SU(N_f)\times
SU(N_f)\times SU(2)_R \times U(1)_\CR $.

The theory has an $N_c-1$ complex dimensional moduli space of vacua,
which are parameterized by the gauge invariant order parameters
\beq
u_k = Tr\langle\phi^k\rangle,~~~~~~ k=2,\ldots,N_c
\comma
\label{invar}
\eeq
$\phi$ being the
scalar field of the $N=2$ chiral multiplet.
The moduli space of vacua corresponds to $\phi$
satisfying a D-flatness condition $[\phi,\phi^\dagger]=0$.
Thus, up
to gauge transformation we can take
\beq
\langle \phi \rangle = \sum_{i=1}^{N_c} a_i H_i =
diag[a_1,\ldots,a_{N_c}] \comma
\label{vev}
\eeq
with $H_i$ being the generators of the Cartan sub-algebra of $U(N_c)$
and
\beq
\sum_{i=1}^{N_c} a_i = 0 \stop
\eeq

At weak coupling we have
\beq
u_k = \sum_{i=1}^{N_c} a_i^k \stop
\label{uka}
\eeq

Alternative gauge invariant order parameters which will be
useful for $N_c > 3$ are defined at the classical level as the
symmetric polynomials $s_k$ in $a_i$
\beq
s_k = (-)^k \sum_{i_1<\cdots<i_k} a_{i_1}\cdots a_{i_k} ,~~~~~~ k=2,\ldots,N_c
\comma
\label{symm}
\eeq

Using (\ref{uka}) and (\ref{symm}) the sets $s_k$ and $u_k$ are
related by Newton's formula \cite{Alon}
\beq
ks_k + \sum_{i=1}^k s_{k-i}u_i = 0 ,~~~~~~ k=\{1,2,\ldots\} \comma
\label{Newton}
\eeq
where $s_0=1,s_1=u_1=0$.
This relation serves as a definition of $s_k$ at the quantum level.

At a generic point the expectation values of $\phi$ break
the gauge symmetry to $U(1)^{N_c-1}$ and a low-energy effective Lagrangian can
be written in terms of multiplets $(A_i,W_i)$, $i=1,\ldots,N_c$,
where $\sum_i A_i=0$.

The $N=2$ effective Lagrangian
takes the form
\beq
\CL_{eff}=Im{\frac{1}{4\pi}}\[\int d^4\theta\partial_i\CF(A)\bar
 A^i+{1\over2}\int
d^2\theta\partial_i\partial_j\CF(A)W^iW^j\] \comma
\label{oneloop}
\eeq
where $\CF$ is a holomorphic pre-potential.

Classically,
\beq
\CF_{cl}(A)={\tau\over2}\sum_{i=1}^{N_c}\(A_i-{1\over N_c}
\sum_{j=1}^{N_c}A_j\)^2 \comma
\label{clf}
\eeq
where $\tau=N_c\({\theta\over2\pi}+{4\pi i\over g^2}\).$
The one loop correction for this pre-potential $\CF$ is given by
\beq
\CF_1=i\frac{2N_c-N_f}{8\pi}\sum_{i<j}(A_i-A_j)^2
\log{(A_i-A_j)^2\over\Lambda^2}
\comma
\label{F1}
\eeq
and $N=2$ super-symmetry
implies that there are no perturbative corrections beyond one loop.
There are, however, non-perturbative instanton corrections.

The classical moduli space is described by the family of genus $g=N_c-1$
hyper-elliptic curves  \cite{Shimon,Alon}
\footnote{Note that the curve (\ref{cl1}) differs from
$y^2 = (x^{N_c}-\sum_{i=2}^{N_c}
\frac{u_i}{i} x^{N_c-i})^2$ for $N_c>3$.}
\beq
y^2={\cal C}^2_{N_c}(x) =
\prod_{i=1}^{N_c}(x-a_i)^2  \comma
\label{cl1}
\eeq
and in terms of the symmetric functions $s_i$
\beq
{\cal C}_{N_c}(x) = x^{N_c}+\sum_{i=2}^{N_c}
s_i x^{N_c-i} \stop
\label{cl}
\eeq
In (\ref{cl1}) $y$ is a double cover of the $x$
plane branched at $N_c$ points
corresponding to the roots $a_i$.
The singularities of the classical curve correspond to cases where the gauge
group is spontaneously broken to non-Abelian subgroups of $SU(N_c)$ rather than
to $U(1)^{N_c-1}$.
Since each root has multiplicity greater than one in (\ref{cl1}),
the classical curve is always singular. Physically,
this is related to the
 fact that the dynamical scale of the theory vanishes and
the states with mass proportional to the scale become massless.

We will construct the quantum moduli spaces of vacua
also as families of
hyper-elliptic curves
\beq
y^2 = \prod_{i=1}^{2g + 2}(x-e_i) \comma
\label{hyper}
\eeq
where the roots $e_i$ are functions of the quark masses $m_i$,
the dynamically generated
scale $\Lambda$
\footnote{When $N_f=2N_c$, the role of the dynamical scale $\Lambda$
is played by a modular form of the appropriate modular group.}
and the
gauge invariant order parameters
$u_k$ or $s_k$, with (\ref{cl1}) being their classical limit.

\subsection{Dyon spectrum and duality}
The spectrum of $N=2$ supersymmetric QCD includes
particles in the ``small" representation of the $N=2$ algebra,
the so called
BPS-saturated states. These are electrically
and magnetically charged particles
with masses
\beq
M^2=2|Z|^2 \comma
\label{MZ}
\eeq
where $Z$ is a central extension of the $N=2$ supersymmetry
algebra and is a linear combination of conserved charges \cite{wo}.
It reads \cite{sw2}
\beq
Z = \sum_{i=1}^{N_c}[n_e^i a^i +n_m^i a_D^i] +\sum_{i=1}^{N_f}S_i m_i \comma
\label{BPS}
\eeq
where $n_e,n_m$ are the electric and magnetic charges
satisfying
$\sum_{i=1}^{N_c}n_e^i=0$,
$\sum_{i=1}^{N_c}n_m^i=0$.
$S_i$ are the $U(1)$ charges corresponding to additional symmetries
that may exist  when the global symmetry is
explicitly broken by non-zero masses.
$m_i$ are the bare masses of the hyper-multiplets
and  $a,a_D$ correspond to the vacuum expectation values of the
scalar component of the chiral super-field $A$ and its dual
\beq
a_D^i=\frac{\partial\CF(a)}{\partial a} \stop
\label{ad}
\eeq

Mathematically,
$a,a_D$ are sections of a certain bundle over the moduli space
of vacua.
When all the masses $m_i$ are zero, $a,a_D$ are sections
of an $Sp(2g,Z)$ bundle, thus their monodromies upon traversing a
closed cycle in the moduli space are elements of
$Sp(2g,Z)$.
The mass formula (\ref{MZ}) is
$Sp(2g,Z)$ invariant, reflecting a duality property of the theory.
When the masses do not vanish, constant shifts are
allowed transformations in addition to the above $Sp(2g,Z)$
structure \cite{sw1,sw2}. Thus, the monodromy transformation
takes the form
\beq
\pmatrix{a_D \cr a}\rightarrow M\pmatrix{a_D \cr a}+c,
{}~~~~~~~M \in Sp(2g,Z)\comma
\label{mondef}
\eeq
where $c$ is independent of the moduli parameters
and depends on the masses of the quarks.

$a,a_D$ can be written as periods of a meromorphic one form $\lambda$
on the curve describing the space of vacua
\beq
a_D^i=\oint_{\alpha_i}\lambda \comma~~~~~~
a^i=\oint_{\beta_i}\lambda \comma
\label{aad}
\eeq
where $\alpha_i,\beta_j$ form a basis of homology cycles on the curve.
In order to determine $\lambda$ one assumes that
\beqar
\frac{\partial a_D^i}{\partial s_k}
& \propto &\oint_{\alpha_i}x^{N_c-k}\frac{d x}{y}
 ~~~~~~k=2,\ldots,N_c\nonumber\\
\frac{\partial a^i}{\partial s_k}
& \propto &\oint_{\beta_i}x^{N_c-k}\frac{d x}{y}~~~~~~k=2,\ldots,N_c  \comma
\label{aap}
\eeqar
with  $x^{N_c-k}\frac{d x}{y}~~~k=2,\ldots,N_c$
being a basis of holomorphic one forms on the curve.
The exact proportionality factor in (\ref{aap}) is determined
by matching $a^i$ to the weak coupling relations (\ref{uka})
or (\ref{symm}). It is a constant since we want to avoid unwanted zeros or
poles.
Its value, as will be calculated in section six,
is ${1\over2\pi i}$.

Note that (\ref{aad}) and (\ref{aap}) yield, up to an exact form,
\beq
\frac{d \lambda}{d s_k} \propto x^{N_c-k}\frac{d x}{y}
{}~~~~~~k=2,\ldots,N_c\stop
\label{lambda}
\eeq

When the bare masses are zero the residues of $\lambda$
vanish, thus ensuring that
 $a,a_D$ are invariant under deformations of the cycle of integration
even across poles of $\lambda$.
For non zero masses the residues take the form
\beq
2 \pi i \res(\lambda) = \sum_in_im_i ~~~~~~ n_i \in {1\over2}Z \comma
\label{res}
\eeq
thus allowing jumps in $a,a_D$, as defined in (\ref{aad}),
when crossing poles of $\lambda$ compatible with the mass formula
(\ref{BPS}).

The matrix of low energy coupling constants, $\tau$, is given by
\beq
\tau_{ij}(a)= \frac{\partial a_D^i}{\partial a^j} \stop
\label{pmatrix}
\eeq
By virtue of (\ref{aap}), $\tau_{ij}$ is the period matrix of the curve
describing the quantum moduli space.
The metric on quantum moduli space reads
\beq
(d s)^2 = Im~ d a_D^i d\bar{a}^i
\comma
\label{metric}
\eeq
is invariant under the transformation (\ref{mondef}),
and is positive definite.

\subsection{The singularity, global symmetry and monodromy structures}

As discussed in section 2.1, the $N_c-1$ dimensional
moduli space of vacua ${\cal M}_{N_c}$ is parameterized
by gauge invariant order parameters such as
$s_k$ of equation (\ref{symm}).
The singular locus of the family of curves (\ref{hyper})
describing the moduli space
is the codimension one variety defined as the
vanishing locus of the discriminant
\footnote{When the coefficient of the highest order monomial $x^n$
in the polynomial is
$a \neq 1$ the discriminant (\ref{var}) is modified by a pre-factor
$a^{2n-2}$.}
\beq
\Delta[s_k] \equiv \prod_{i<j}(e_i-e_j)^2 = 0 \stop
\label{var}
\eeq
Along this variety additional massless states appear in the spectrum and the
effective low energy description (\ref{oneloop})
is not valid.

In the semi-classical limit $\Lambda\rightarrow 0$ the discriminant
factorizes
\beq
\Delta[\Lambda\rightarrow 0] = \Lambda^{N_c(2N_c-N_f)}
\Delta_{N_c}^2\Delta_{N_f,N_c} \comma
\label{Del}
\eeq
where
\beq
\Delta_{N_c} = \prod_{i < j}^{N_c}(a_i-a_j)^2 \comma
\label{D0}
\eeq
and
\beq
\Delta_{N_f,N_c} = \prod_{j=1}^{N_f}\(\sum_{i=0}^{N_c}s_i(-m_j)^{N_c-i}\) \stop
\label{D1}
\eeq
The zero locus of $\Delta_{N_c}$ corresponds to singularities
where classically the $SU(N_c)$ group is not spontaneously broken
to $U(1)^{N_c-1}$ but rather to a larger subgroup (In particular,
the gauge group is not broken at all at the origin of the moduli space).
The zero locus of $\Delta_{N_f,N_c}$ defines a complex codimension one
variety in the moduli space where a quark becomes massless
classically. In order to see this note that
the singular locus corresponding to a massless quark,
$a_i + m_j =0,~~i = 1,\ldots,N_c,~~j = 1,\ldots,N_f$,
is a codimension
one variety defined by
\beq
\prod_{i=1}^{N_c}(a_i+m_j)\equiv
\sum_{k=0}^{N_c}s_k(-m_j)^{N_c-k} = 0
\comma
\qquad j=1,\ldots,N_f\comma
\label{classvar}
\eeq
where (\ref{classvar}) is derived by reading the classical mass of the quarks
from the super-potential (\ref{superpot})
and using the classical curve (\ref{cl1})
\footnote{In order not to carry factors of
$\sqrt{2}$ we re-scale the mass appearing in (\ref{superpot}) by
$m\rightarrow\sqrt{2}m$.}.
The product up to $N_f$ in (\ref{D1}) is a consequence of having $N_f$
different flavors.
The power of $\Lambda$ in (\ref{Del}) follows from instanton contribution and
from the fact that classically $N_c$ roots degenerate.
The factorization (\ref{Del})
holds quantum mechanically with the symmetric
functions being modified by quantum corrections.

While singularities of the curves
are associated with
the appearance of additional massless states in the spectrum,
the order of vanishing of the discriminant at a point in the
moduli space corresponding
to a singularity of the curves is generically the number of
codimension one varieties intersecting at the point, which in turn
is the number
of
massless states at that point.
These states  belong to a representation of the global symmetry
group and the order of vanishing of the discriminant should be related to
its dimension.
Thus, compatibility between
the global symmetry and the singularity structure
imposes constraints on the latter.

Note, however,
that different types of singularities of the curves
are associated with different physics.
As discussed in \cite{DS}, when more than
two branch points of the curve coincide the massless states
at the corresponding point in the moduli space are not mutually local.
The singular points corresponding to $N=1$ vacua, upon adding a mass
term as perturbation, are associated with curves of degree $2n$ with
$n-1$ of pairs of identical branch points.

We shall prove in the sequel that the moduli space of vacua
for $N_f < N_c$ is fully determined by $R$-symmetry, instanton contributions
and
classical singularity without any need for constraints
coming from global symmetry considerations, thus
consider the cases
$N_f \geq N_c$.
Let us present an observation that we have about the relation between
the global symmetries and the singularity structure,
for which we do not have a proof.
This observation will be used by us in order to suggest a complete
determination of the curves for $N_f \geq N_c$.

Consider the theories with $N_c$ massless multiplets and
$N_f - N_c$ massive multiplets with the same bare mass $m$.
Denote by ${\cal L}$ the variety defined by
(\ref{var}).
Let $l$ be a complex line in the moduli space defined
by $s_i=0$ for all but one modulus, and consider the intersection
of ${\cal L}$ and $l$.
${\cal L}\cap l$ exhibits a singularity
structure with three singular points of multiplicity
$(N_c,N_c,N_f-N_c)$.
This is probably related to a global symmetry group
$SU(N_c)\times SU(N_c)\times SU(N_f-N_c)$.

The singularity structure should also reflect itself in the
monodromy structure.
The monodromy group of the family of curves (in the massless
case)
is the homomorphic image of the fundamental group
$\Pi_1({\cal M}_{N_c} - {\cal L})$ in $Sp(2N_c -2,Z)$.
Physically, the monodromy matrices (in the strong coupling regime)
specify the electric and magnetic
quantum numbers of the massless dyon associated with the
singularity. These electric and magnetic charges are the
left eigenvectors of the
monodromy matrix with eigenvalue one.

A non trivial monodromy exists also at
infinity in the moduli space corresponding
to the semi-classical regime of the theory.
This monodromy is not associated with additional massless states
but rather with the logarithm in (\ref{oneloop}).
This monodromy will be computed in section six
for arbitrary $N_c$ and $N_f$, as a check on the curves.

\subsection{Principles for the construction of the moduli spaces}

In the following we summarize the
principles which will be used by us in the construction of the
families of hyper-elliptic curves describing the quantum
moduli spaces of vacua:

\begin{enumerate}
\item {{\bf Symmetry} : The curves are invariant under $R$ charge
transformation
\beq
O \rightarrow \exp\[\frac{2 \pi i R(O)}{4(2N_c -N_f)}\] O \comma
\label{Rs}
\eeq
where $R(O)$ is the $R$ charge of $O$ and $O$ refers to the
building blocks of the curves, $y,x,m_i,\Lambda,s_k$.
We have
\begin{center}
\label{Rc}
\begin{tabular}{|c||c|c|c|c|c|c|}
\hline
$O$ & $y$ & $x$ & $m_i$ & $\Lambda$ & $u_k$ & $s_k$ \\ \hline
$R(O)$ & $2N_c$ & $2$ & $2$ & $2$ & $2k$ & $2k$ \\ \hline
\end{tabular}
\end{center}
}

\item {{\bf Singularity structure} : As discussed in section
2.3, singularities of a curve
describing a quantum moduli space
are associated with the zero locus of the corresponding
discriminant and are physically interpreted as
the appearance of additional massless states in the spectrum,
where the order of vanishing of the discriminant at a point corresponding
to a singularity of the curve is generically the number of
massless states at that point.
These states  belong to a representation of the global symmetry
group and the order of vanishing of the discriminant is
its dimension. Thus, global symmetry imposes constraints on the
singularity structure.}

\item {{\bf Instanton corrections} : The one instanton process contribution
to the curves for $N_f <2 N_c$ takes the form \cite{nati}
\beq
\Lambda^{2N_c-N_f} \stop
\label{instant}
\eeq
}

\item {{\bf Integration of a massive quark}:
Sending a quark mass $m\rightarrow \infty$ in $N_f < 2N_c$
theories,
and sending the scale $\Lambda_{N_f} \rightarrow 0$ such that
\footnote{In \cite{FP} this renormalization scheme was identified as the
$\overline{DR}$ scheme.}
\beq
\Lambda_{N_f-1}^{2N_c-N_f+1} = m\Lambda_{N_f}^{2N_c-N_f}
\comma
\eeq
is fixed,
we reduce the number of flavors from
$N_f$ to $N_f-1$ and we require compatibility of the curves.
When $N_f=2N_c$ the appropriate matching condition reads
\beq
\Lambda_{2N_c-1} = 64m q \comma
\label{ql}
\eeq}
where $q \equiv e^{2\pi i\tau}$ and
the constant $64$ is a choice of a renormalization scheme.
\item {{\bf Classical limit}: The curves should exhibit the singularity
structure (\ref{Del})-(\ref{D1})
at the classical level
$\Lambda \rightarrow 0$.
}
\item {{\bf Residues} :
As discussed in section 2.2,
the meromorphic one form on the curve $\lambda$ in (\ref{aad})
may have poles but its residues are restricted
by (\ref{res}).
This, as we shall see, leads to powerful constraints on the
structure of the curves.
}
\end{enumerate}

Our aim is to construct the hyper-elliptic curves describing the
quantum moduli spaces of $N=2$ QCD with $N_c$ colors and $N_f$
flavors in the fundamental representation
of the gauge group. For this, the form of the hyper-elliptic curves introduced
in \cite{Shimon,Alon} is the appropriate framework.
The general strategy for constructing the curves will be to first
restrict the
possible terms to those compatible with $R$ charge symmetry and the
form of the instanton corrections. We then impose the proper
classical singularity structure and use residue calculations.
Finally, we make use of the global symmetry and its implications
on the singularity structure.
Monodromy calculations will be used as a consistency check.

Since the one loop beta function of the theory is proportional to $2N_c-N_f$
(higher loop corrections to it vanish) we limit ourselves to $N_f<2N_c$ where
 the
theory is asymptotically free and to $N_f=2N_c$ where the
(massless) theory is scale
invariant.

\section{The quantum moduli space for  $N_f < N_c$}

\subsection{The general case}
For $N_f < N_c$ there exist flat directions along which the
$SU(N_c)$ gauge group is generically broken to $U(1)^{N_c-1}$ and
the theory is in a coulomb phase: There is a collection of $N_c-1$ massless
photons in the spectrum.
In these regions, as we shall argue, $R$-symmetry, classical
singularity and the form of the instanton corrections are
sufficient in order to fully determine the quantum moduli spaces.
First recall the curve for $N_f=0$. It reads
\cite{Shimon,Alon}
\beq
y^2 = {\cal C}^2_{N_c}(x) - \Lambda_0^{2N_c} \comma
\label{0n}
\eeq
where ${\cal C}_{N_c}(x)$ is given by (\ref{cl}).
\newline
When $N_f$ flavors in the fundamental representation of
the gauge group are present such that $N_f <N_c$ we have
the following claim:
\newline
{\bf Claim}: The curve describing
the quantum moduli space with
$N_f<N_c$ flavors is:
\beq
y^2 =  {\cal C}_{N_c}(x)^2- \Lambda_{N_f}^{2N_c-N_f}
\prod_{i=1}^{N_f}
(x+m_i) \stop
\label{gm}
\eeq
{\bf Proof}:
$R$-symmetry, the form of instanton corrections together with the classical
singularity of the gauge group imply that the most general curve is
\beq
y^2 =  {\cal C}_{N_c}(x)^2-
\Lambda_{N_f}^{2N_c-N_f}G(x,m_i) \equiv P \comma
\label{gm1}
\eeq
where $G(x,m_i)$ is a polynomial of degree $N_f$ in $x$.
A priori, $G$  may also depend on the moduli
$s_k$. In the sequel it will be shown that in fact it is independent
of the moduli.
The second term in (\ref{gm1}) is the quantum correction  to the
classical curve (\ref{cl1}). Note that there are
only one instanton contributions to the curve (\ref{gm1}) due to
the $R$ charge restriction.
The polynomial $G(x,m_i)$ is determined
by requiring that the discriminant
of the polynomial $P$
in (\ref{gm1}) has the right classical
limit (\ref{Del}).
In order to do this we need to evaluate first the classical
limit of the discriminant.
Note that the discriminant of a polynomial can be evaluated up to a
multiplicative constant by
\footnote{The discriminant of a degree $n$
curve is a polynomial in its roots of degree $n(n-1)$ that
vanishes when any two roots of the curve coincide. It is easy to
see that (\ref{discdef}) satisfies these.}
\beq
\Delta[P] = \prod_{x_i \in S}P(x_i) \comma
\label{discdef}
\eeq
where $x_i \in S$ are the critical points of $P$,
$\partial_x P(x_i) = 0$.
Differentiating the polynomial in (\ref{gm1}) with respect
to $x$ and equating to zero we have
\beq
P' = 2{\cal C}_{N_c}(x)
{\cal C}_{N_c}(x)'-
\varepsilon G(x,m_i)' =0 \comma
\label{gm1d}
\eeq
where we denoted $\varepsilon =  \Lambda_{N_f}^{2N_c-N_f}$.
The roots of equation (\ref{gm1d}) are $\{r_i=a_i +
\varepsilon~~ corrections \}$ and
$\{s_i = b_i + \varepsilon~~ corrections \}$ where $a_i$ and $b_i$ are
the roots of ${\cal C}_{N_c}(x)$ and ${\cal C}_{N_c}(x)'$, respectively.
According to (\ref{discdef})
\beq
\Delta[P] = \prod_iP(r_i)\prod_jP(s_j) \stop
\label{val}
\eeq
In order to analyze the classical limit
we have to evaluate (\ref{val}) as $\varepsilon \rightarrow 0$.
Consider the first product in (\ref{val}) and let us prove that
to lowest order in $\varepsilon$
\beq
\prod_iP(r_i) = \varepsilon^{N_c}\prod_i G(a_i,m_j)
\comma
\label{val1}
\eeq
which is the contribution of the  $\varepsilon G$ term
in $P$.
We have to show that the  ${\cal C}_{N_c}(x)$ contribution
is of higher order in $\varepsilon$.
Suppose $a_i$ is a root of ${\cal C}_{N_c}(x)$ of multiplicity $n$, then
it follows from  (\ref{gm1d}) that
\beq
r_i = a_i + c_i \varepsilon^{\frac{1}{2n-1}} +\ldots \stop
\label{reps}
\eeq
Thus the contribution of ${\cal C}_{N_c}(r_i)^2$ is of order
$\frac{2n}{2n-1} > 1$ in $\varepsilon$ which is higher than
the contribution of the $\varepsilon G$ term.
In order to get the lowest order in $\varepsilon$
we evaluate the
second product in (\ref{val}) at the roots $b_i$, which yields
using (\ref{discdef})
\beq
\prod_jP(b_j) = \Delta[{\cal C}_{N_c}]^2  \comma
\label{val2}
\eeq
where  $\Delta[{\cal C}_{N_c}]$, given by (\ref{D0}) is the discriminant
of the classical curve (\ref{cl1}). Thus,
(\ref{val1}) and (\ref{val2}) yield
\beq
\Delta[P] = \varepsilon^{N_c}\Delta[{\cal C}_{N_c}]^2
\prod_i G(a_i,m_j) \stop
\label{val3}
\eeq
A comparison of (\ref{val3}) to (\ref{Del}),(\ref{D0})
and (\ref{D1}) yields
\beq
G(x,m_i) =
\prod_{i=1}^{N_f}
(x+m_i) \comma
\label{G}
\eeq
which completes the proof.

The meromorphic one-form $\lambda$, satisfying
(\ref{lambda}), takes the form
reads
\beq
\lambda = \frac{x dx}{2\pi iy}\( \frac{{\cal C}_{N_c}
G'}{2G}-{\cal C}_{N_c}' \) \comma
\label{lam}
\eeq
for $N_f<N_c$.
The fact that the residues of $\lambda$ satisfy (\ref{res}) is
a corollary of the residue calculation that will be presented
in the next section.
This provides a consistency check on our result.

\subsection{Examples}
{\bf $N_c=2$}:
Denote the gauge invariant order parameter $\frac{u_2}{2}$
of (\ref{invar}) by $u$.\newline
{\bf $N_f=0$}: The curve is given by \cite{Shimon,Alon}
\beq
y^2 = (x^2-u)^2 - \Lambda_0^4 \stop
\label{0}
\eeq
Let us, as an example, compute the periods and
the proportionality constant of (\ref{aap}) in this case.
We will show in section six that the result is in fact general.
The roots of (\ref{0}) are  $\pm x_1=
\pm \sqrt{u+\Lambda_0^2}$ and $\pm x_2=\pm \sqrt{u-\Lambda_0^2}$.

Define the $\alpha$ and $\beta$ cycles
as the contours from $-x_1$ to $x_1$ and from $x_1$ to $x_2$
respectively and back counterclockwise.
The corresponding periods of the curve read
\beqar
\omega_1 = 2\int_{-x_1}^{x_1} \frac{{\rm d}x}
{\sqrt{(x^2-u)^2 - \Lambda_0^4}} = \frac{2\pi}{x_2}
F\(\frac{1}{2},\frac{1}{2},1,\frac{x_1^2}{x_2^2}\) \comma \nonumber\\
\omega_2 = 2\int_{x_1}^{x_2} \frac{{\rm d}x}
{\sqrt{(x^2-u)^2 - \Lambda_0^4}} = -\frac{2\pi i}{x_1+x_2}
F\(\frac{1}{2},\frac{1}{2},1,\frac{(x_1-x_2)^2}{(x_1+x_2)^2}\) \comma
\label{period}
\eeqar
where $F$ is the hyper-geometric
function. The behavior of the periods as $u\rightarrow \infty$ is
\beq
\omega_1 \rightarrow {2\over\sqrt{u}}\log\(\frac{u}{2 \Lambda_0^2}\),~~~~~
\omega_2 \rightarrow -\frac{\pi i}{\sqrt{u}} \stop
\label{lim}
\eeq
Thus,
\beqar
a \simeq -2 \pi i \sqrt{u} \comma \nonumber\\
a_D \simeq i \frac{4}{\pi} a \log(a) \stop
\label{aad1}
\eeqar
Since at weak coupling $a = \sqrt{u}$ (\ref{uka})
we have
\beqar
\frac{{\rm d}a_D}{{\rm d} u}
=-\frac{1}{2 \pi i}\oint_{\alpha}\frac{d x}{y} \comma~~~~~~
\frac{{\rm d}a}{{\rm d} u}
=-\frac{1}{2 \pi i}\oint_{\beta}\frac{d x}{y} \stop
\eeqar
\newline
{\bf $N_f=1$}: As in the proof of (\ref{gm})
in order to determine
the curve we make use of $R$-symmetry, the form of
instanton corrections and the scaling and
classical limits.
The curve reads
\beq
y^2 = (x^2-u)^2 - \Lambda_1^3(x+m) \stop
\label{11}
\eeq
The  coefficient of $m\Lambda_1^3$ in (\ref{11})
is determined by integrating the massive quark and
requiring compatibility with (\ref{0}), while that of $x\Lambda_1^3$ is
fixed by the requirement for having a
singularity at $u=m^2$ at the
classical limit $\Lambda_1\rightarrow 0$
corresponding to a massless quark.
The discriminant of the curve (\ref{11})
\beq
\Delta_{2,1}=-\Lambda_1^6(256u^3-256u^2m^2-288um\Lambda_1^3+256m^3\Lambda_1^3
+27\Lambda_1^6)
\comma
\eeq
has a novel symmetry: It
is invariant under the
transformation $u\rightarrow m \Lambda_1, m\rightarrow \frac{u}{\Lambda_1}$.
This implies that the corresponding curves
are equivalent : There exists an $SL(2,C)$ transformation taking one
curve to the other.
That means that all the physical quantities associated with the
quantum moduli space of vacua such as
the BPS spectrum are invariant under this transformation.
In particular, the massive theory with mass $m$ at the origin of the moduli
space $u=0$ is equivalent to a massless theory at $u=m\Lambda_1$ on the
moduli space.
We expect such relations to hold for other theories.
\newline
{\bf $N_c=3$}:
Denote the gauge invariant order parameter $\frac{u_3}{3}$
of (\ref{invar}) by $v$.\newline
{\bf $N_f=0$}: The curve takes the form \cite{Shimon,Alon}
\beq
y^2 = (x^3-ux - v)^2 - \Lambda_0^6 \stop
\label{03}
\eeq
\newline
{\bf $N_f=1$}: The curve reads
\beq
y^2 = (x^3-ux - v)^2 - \Lambda_1^5 (x+m) \stop
\label{113}
\eeq
The coefficient of $m\Lambda_1^5$ is determined by
integrating a massive quark and requiring compatibility with the
$N_f=0$ case.
The coefficient $a$ of $x\Lambda_1^5$ is determined by
imposing the expected classical singularity, as we will show now.
The singular complex line associated with a classical massless quark is
the variety defined by
\beq
\prod_{i=1}^3(a_i+m) \equiv v - mu+m^3 = 0 \stop
\label{sing}
\eeq
The lowest order term in $\Lambda_1$ of the discriminant of the curve
(\ref{113}) reads
\beq
(4u^3-27v^2)^2(m^3-a^2um + a^3v) \comma
\label{disc}
\eeq
where we factored out $64 \Lambda_1^{15}$.
The zero locus of the first term
corresponds to singularities where classically the $SU(3)$ group
is not spontaneously broken to $U(1)^3$ but rather to the
subgroup $SU(2)\times U(1)$,
and to the case where it is not broken at all at the origin $u=v=0$.

The zero locus of the second term is a complex line in the
moduli space were a quark becomes massless classically.
Comparison with (\ref{sing}) fixes the value of the coefficient to one.
\newline
{\bf $N_f=2$}:
The curve reads
\beq
y^2 = (x^3-ux - v)^2 - \Lambda_2^4 (x+m_1)(x+m_2) \stop
\label{nf2}
\eeq
The terms $a x^2 + b u$ which are allowed by $R$-symmetry and
instanton corrections are excluded by
imposing the classical singularity structure.

The discriminant of (\ref{nf2}) for equal masses $m_1=m_2=m$ reads
\beq
\Delta_{3,2} = 64\Lambda_2^{12}\left(v-mu+m^3\right) ^2
\Delta_+ \Delta_- \comma
\label{disnf2}
\eeq
where
\beq
\Delta_{\pm}=\[27(v \pm m\Lambda_2^2)^2-4(u \pm \Lambda_2^2)^3\]
\stop
\eeq
The multiplicity two in the discriminant (\ref{disnf2})
is a reflection of the underlying $SU(2)$ global
flavor symmetry: The massless states along this line are
in the fundamental representation of this group.
As $\Lambda \rightarrow 0$, (\ref{disnf2})
provides an example to the classical singularity structure
(\ref{Del}).

\section{The quantum moduli space for  $N_f = N_c$}

\subsection{The general case}

The most general curve consistent with $R$-symmetry,
instanton corrections and classical singularity is
\beq
y^2 =  {\cal C}^2_{N_c}(x)- \Lambda^{N_c}\( \prod_{i=1}^{N_f}
(x+m_i) + a{\cal C}_{N_c}(x)\) + b \Lambda^{2N_c} \comma
\label{ncnf}
\eeq
where $a$ and $b$ are constant coefficients.
Since the one instanton correction $\Lambda^{N_c}$ has $R$ charge
$N_c$, the curve (\ref{ncnf}) gets contributions also from a two
instanton process of the form $\Lambda^{2N_c}$.
The structure of the correction  $\Lambda^{N_c}{\cal C}_{N_c}(x)$
in (\ref{ncnf}) is determined such that it vanishes for $x=a_i$
as required, via the analysis of the previous section,
by comparing
(\ref{val3}) with (\ref{D1}).

The curve (\ref{ncnf}) can be written in a form suitable
for generalization to $N_f>N_c$:
\beq
y^2 =  F^2_{N_c} - H_{N_c} \comma
\label{ncnf1}
\eeq
where
\beqar
F_{N_c} & = &  {\cal C}_{N_c} + \alpha \Lambda^{N_c} \nonumber\\
H_{N_c} & = &
\Lambda^{N_c}\( \prod_{i=1}^{N_f}(x+m_i)  + \beta \Lambda^{N_c}\) \comma
\label{ncnf3}
\eeqar
with $\alpha$ and $\beta$ constants.

The meromorphic one-form $\lambda$ (\ref{lambda}) reads in this case
\beq
\lambda = \frac{x dx}{2\pi iy}\( \frac{F_{N_c} H_{N_c}'}{2H_{N_c}}-F_{N_c}' \)
 \stop
\label{lamn}
\eeq
The residue formula (\ref{res}) can be used in order to determine the
coefficient $\beta$, as we will show now.
This will be an example of a powerful constraint which will be much used
in the sequel.

Let us consider the case of equal bare masses $m_i=m$. The zeros of
$H_{N_c}$ in (\ref{ncnf3}) read
\beq
x_i = -m + e^{\frac{2 \pi i}{N_f}} \beta^{\frac{1}{N_f}}\Lambda
\comma~~~~~~i=1,...,N_f-1 \stop
\label{roots}
\eeq
The residue of $\lambda$ in (\ref{lamn})
at the root $x_0$ of (\ref{roots}) is
\beq
2\pi i\res_{x=x_0}(\lambda) = \frac{m-\beta}{2} \comma
\label{res1}
\eeq
thus (\ref{res}) implies $\beta =0$.
The residues of $\lambda$ at zeros of $y$ vanish.
This is easily seen by differentiating (\ref{ncnf1}) with respect to $x$,
which together with (\ref{ncnf1}) at $y=0$ yield
\beq
\(\frac{F_{N_c}
H_{N_c}'}{2H_{N_c}}-F_{N_c}' \)_{y=0} =0 \stop
\eeq
Thus, we verified that the residue formula (\ref{res}) is satisfied
completely by $\lambda$ of (\ref{lamn}).
Note that we are still left with one undetermined constant
coefficient $\alpha$.
When $N_c=2$ it is fixed by
imposing a $Z_2$ symmetry on the singular locus of the
curve, as will be shown in section 4.2.

Compatibility with the hyper-elliptic curves for $N_c>N_f$ which will
be discussed in the next section implies that $\alpha =\frac{1}{4}$
for $N_c>2$, however, we do not have a full proof for that.

Thus, we suggest that the family of curves
describing the quantum moduli space of vacua
for the $N_c=N_f,~N_c>2$ is
\beq
y^2 =  \({\cal C}_{N_c}(x)+ \frac{\Lambda^{N_c}}{4}\)^2
- \Lambda^{N_c} \prod_{i=1}^{N_f}
(x+m_i) \stop
\label{nn}
\eeq

\subsection{Examples}
{\bf $N_c=N_f=2$}:
The curve takes the form
\beq
y^2 =  \(x^2-u+{\Lambda_2^2\over8}\)^2- \Lambda_2^4(x+m_1)(x+m_2) \stop
\label{21n}
\eeq

$R$-symmetry, instanton corrections, classical
singularity and scaling limit, leave us with a two parameter family of curves
 given by
\beq
y^2=(x^2-u+a\Lambda_2^2)^2-\Lambda_2^2(x+m_1)(x+m_2)-b\Lambda_2^4
\stop
\eeq
The singularity structure of the massless theory:
Two singular points of multiplicity two, leads to $b=0$.
The requirement that the singular points be located in a $Z_2$
symmetric
form in the moduli space leads to $a={1\over8}$.
There are two cases where an underlying global $SU(2)$ structure
appears.
First, when $m_2=m_1=m$
the discriminant of the curve takes the form
\beq
{\Lambda_2^4\over16}(8u-8m\Lambda_2+\Lambda_2^2)(8u+8m\Lambda_2+\Lambda_2^2)
(8u-8m_1^2-\Lambda_2^2)^2 \stop
\eeq
The multiplicity two reflects the fact that the massless states along
the corresponding singular line are in the fundamental representation of
the flavor symmetry group.
Second, when $-m_2=m_1=m$
the discriminant takes the form
\beq
{\Lambda_2^4\over16}(64u^2-16u\Lambda_2^2+64m^2\Lambda_2^2+\Lambda_2^4)
(8u-8m_1^2-\Lambda_2^2)^2 \comma
\eeq
with the multiplicity two showing up.
\newline
{\bf $N_c=N_f=3$}:
\newline
The curve takes the form
\beq
y^2=\(x^3-ux-v+{\Lambda_3^3\over4}\)^2-\Lambda_3^3\prod_{i=1}^3\(x+m_i\)
\stop
\eeq
Following the discussion in section 4.1,
we are left with the coefficient of
$\Lambda_3^3$ as the only undetermined parameter.
It is determined to be $\frac{1}{4}$ by matching to the
$N_c=3,N_f=4$ curve upon integration of a
massive quark. The latter curve will be
determined in the subsequent section.

\section{The quantum moduli space for $N_f > N_c$}

When the number of flavors is increased the curves describing the
quantum moduli spaces may get contributions from higher multi-instanton
processes. This increases the number of terms
in the polynomials describing the curves that should be
determined.
In this section we study the general $N_f>N_c$ cases.
We parameterize the curves and determine there structure up to a certain
number of unknown constant coefficients.
Global symmetry considerations suggest a fixed value for these
constants, but we do not have a solid proof for that.

\subsection{$N_f=N_c+1$}
As a preliminary to the general case, let us add another
flavor to the $N_f=N_c$ case which has been analyzed in the
previous section.
Now the curve is parameterized by two unknown constants.
The curve has the form
\beqar
y^2 =  F^2_{N_c} - H_{N_f} \comma~~~
F_{N_c}  =   {\cal C}_{N_c} + \Lambda^{N_c-1}(ax+bt_1(m)) \nonumber\\
H_{N_f}  =   \Lambda^{N_c-1} \prod_{i=1}^{N_f}(x+m_i) \stop
\label{ncnfp1}
\eeqar
Where $t_k$ denotes the symmetric function in $m_i$ of order $k$
\beq
t_k(m)=\sum_{i_1<,\cdots,<i_k}m_{i_1}\cdots m_{i_k}.
\eeq
The coefficient $b$ is determined to be ${1\over4}$ by matching to the
$N_f=N_c$
case upon integrating a massive quark.
The coefficient $a$ takes the value ${1\over4}$
following the discussion in section 2.3:
We set the masses of all the quarks but one to zero and
consider the complex line $s_i=0 , i\neq N_c$.
We expect the discriminant to have multiplicities
$(N_c,N_c)$ which is indeed the case only for this value of $a$.

The final form of the curve is
\beq
y^2 =
\[{\cal C}_{N_c} + \Lambda^{N_c-1}\({x\over4}+
{t_1(m)\over4}\)\]^2 -
\Lambda^{N_c-1} \prod_{i=1}^{N_f}(x+m_i) \stop
\label{ncnfp2}
\eeq
\subsection{Examples}
{\bf $N_c=2$, $N_f=3$}:
$R$-symmetry, instanton corrections and classical singularity
structure restrict the form of the curves to
\beqar
y^2&=&{{\(x^2-u+\Lambda_3\({\( m_1 + m_2 + m_3\)\over8}+bx\)\)}^2}-
   \Lambda_3\left[(x+m_1)(x+m_2)(x+m_3)\right. \nonumber\\
 &+& a\Lambda_3^3 +
      b\Lambda_3^2
       \left( m_1 + m_2 + m_3 \right)  +
      c\Lambda_3\left( m_1m_2 + m_1m_3 + m_2m_3 \right)
    + d\Lambda_3u\nonumber\\ &+&\left.
      \left( e\Lambda_3^2 + f\Lambda_3\left( m_1 + m_2 + m_3 \right) \right) x
 + g\Lambda_3x^2  \] \stop
 \label{nf3}
\eeqar
Applying the above residue considerations we find that
most of the terms in (\ref{nf3}) vanish, and
imposing a singularity of order four in the massless
case, which follows from the underlying $SO(6)$ symmetry \cite{sw2},
leads to
$b=0$ or $b={1\over4}$.
The $SO(4)$ global symmetry for the flavor
masses $(m,0,0)$ implies a multiplicity structure
$(2,2)$ in the discriminant and leads to $b={1\over4}$.
Thus, the final curves takes the form
\beq
y^2={{\(x^2-u+\Lambda_3\({\( m_1 + m_2 + m_3\)\over8}+{x\over4}\)\)}^2}-
   \Lambda_3(x+m_1)(x+m_2)(x+m_3).
\eeq
\newline
{\bf $N_c=3$, $N_f=4$}:\newline
The curve takes the form
\beq
y^2=\[x^3-ux-v+{\Lambda_4^2\over4}\(x+t_1(m)\)\]^2-\Lambda_4^2
\prod_{i=1}^4(x+m_i) \stop
\eeq
Its derivation is a special case of the discussion in section 5.1.
\subsection{The general case $N_f > N_c$}

The general structure of the family of curves describing the
quantum moduli space of vacua when $N_f>N_c$ is encoded in the
following claim.\\
{\bf Claim}: The curve describing the quantum moduli space with
gauge group $SU(N_c)$ and
$N_f>N_c$ flavors is:
\beq
y^2 = \({\cal C}_{N_c}(x)+\Lambda^{2N_c-N_f}P\)^2-\Lambda^{2N_c-N_f}
\prod_{i=1}^{N_f}(x+m_i)
\label{nfgnc}
\eeq
where $P(x,m_i,\Lambda)$ is a polynomial of degree $N_f-N_c$ in
$x,m_i$ and is {\it independent} of the moduli $s_k$.\\
{\bf Proof}: Consider the most general hyper-elliptic curve.
Using $\partial_k \partial_l\lambda=
\partial_l \partial_k\lambda$ where
$\partial_k \equiv \frac{\partial}{\partial s_k}$ together with
(\ref{lambda}) yields
\beq
x^{-k} \partial_l y^2 =
x^{-l} \partial_k y^2 \stop
\label{eqh}
\eeq
Equation (\ref{eqh})
implies that $y^2$ depends on the moduli
$s_k$ only via ${\cal C}_{N_c}(x)$.
Thus, $y^2(s_k)= y^2({\cal C}_{N_c}(x)).$
Moreover, $R$ charge symmetry implies that only terms
up to quadratic in  ${\cal C}_{N_c}(x)$ can appear:
\beq
y^2 = {\cal C}_{N_c}^2 g_0 +{\cal C}_{N_c}g_1(x,m_i,\Lambda) +
g_2(x,m_i,\Lambda) \comma
\label{sol}
\eeq
where $g_0$ is a polynomial of degree $0$ in $x$, namely a constant
\footnote{As we will discuss in the sequel,
when $N_f=2N_c$, $g_0$ may be
a modular form $g_0(q)$.},
$g_1,g_2$ are polynomials in $x$ of degree $N_c$
and $2N_c$, respectively and are
independent of the moduli $s_k$.
The classical limit fixes $g_0=1$. The curve (\ref{sol}) may be
recast in the form
\beq
y^2 = \({\cal C}_{N_c}+\Lambda^{2N_c-N_f}P\)^2-\Lambda^{2N_c-N_f}
G \comma
\label{gf}
\eeq
with
\beq
G(x,m_i,\Lambda) = \prod_{i=1}^{N_f}(x+m_i) +
 \sum_{k=1}^{\[\frac{2N_c}{2N_c-N_f}\]-1}
\Lambda^{k(2N_c-N_f)}n_k(x,m) \comma
\label{nfgnc1}
\eeq
where we used $R$ charge considerations and the form of instanton
corrections.
The first term in (\ref{nfgnc1})
has been deduced from the structure of the
classical singularity in section $3$.
The other terms are arranged according to the
order of multi-instanton contribution.
$n_k$ is a polynomial of degree $(k+1)N_f-2kN_c$ and
$\[\frac{2N_c}{2N_c-N_f}\]$ denotes its integer value.
$P(x,m_i,\Lambda)$ is a polynomial of degree $N_f$ in $x,m_i$.

We will now use the residue formula
(\ref{res}) in order to determine the form of
$G(x)$ in (\ref{gf}).
The meromorphic one-form $\lambda$ (\ref{lambda}) associated with
(\ref{gf}) is
\beq
\lambda = \frac{x dx}{2 \pi i y}\(\frac{F
G'}{2G}-F'\) \comma
\label{lam1}
\eeq
where, as before, prime denotes derivative with respect to $x$ and
$F= {\cal C}_{N_c}+\Lambda^{2N_c-N_f}P$.

Consider the zeros $x_i$ of $G(x)$ where $y(x_i)\neq 0$ and let
us evaluate $\res(\lambda)$ at these points.
Since $F/y =\pm 1$ at $x_i$ and $G'/G$ equal
the order $n$ of the root $x_i$
we get
\beq
2\pi i\res_{x=x_i}(\lambda)= \pm {n\over2}x_i
\stop
\label{gres}
\eeq
Comparing (\ref{gres}) to (\ref{res}) we conclude that the roots of
$G(x)$ do not receive quantum corrections, thus
$n_k=0$ in (\ref{nfgnc1}). This completes the proof of the claim.

In order to fully construct the curve we still have to determine
the polynomial $P$.
Studies of the structure of the singularities,
compatibility with the global
symmetries and the observation on a relation between them, made in section
2.3,
suggest the form of $P$ and
the curve for $N_f>N_c$ with $N_c>2$:
\beq
y^2=\[{\cal C}_{N_c}+{\Lambda^{2N_c-N_f}\over4}\sum_{i=0}^{N_f-N_c}x^
{N_f-N_c-i}t_i(m)\]^2-\Lambda^{2N_c-N_f}\prod_{i=1}^{N_f}(x+m_i)
\label{general}
\stop
\eeq
Note that (\ref{general}) includes (\ref{gm}) and (\ref{nn}) as special
cases, and that it gets only one and two-instanton contributions.
However,we do not have a proof for this formula.

\subsection{Examples}
{\bf $N_c=3$, $N_f=5$}:
The most general curve consistent with the general structure
(\ref{nfgnc}) reads
\beqar
y^2=\[x^3-ux-v+\Lambda_5(ax^2+x(bt_1(m)+c\Lambda_5)+dt_2(m)+e\Lambda_5t_1(m)
+f\Lambda_5^2)\]^2 \nonumber\\
-\Lambda_5\prod_{i=1}^5(x+m_i) \stop
\eeqar
Let us now imply global symmetry considerations in
order to determine the curve.
Consider the complex line $u=0$ and set the flavor masses
to the values $(m,m,0,0,0)$.
We expect that the discriminant of the curve exhibit multiplicity
structure of
the form (3,3,2).
The values of $e$ and $f$ do not affect the analysis
since they correspond to shifts in the moduli parameter
$v$, and thus can be set to zero for the meantime.
Requiring the above multiplicity structure
implies that $a=b=c={1\over4}$ and $d=0$.
In order to determine the values of $e$ and $f$
consider the complex line $v=0$
with mass values as before.
Requiring the same multiplicity
structure yields the values
$e=f=0$.
Thus, the final form of the curve is
\beq
y^2=\[x^3-ux-v+{\Lambda_5\over4}(x^2+xt_1(m)+t_2(m))\]^2-\Lambda_5
\prod_{i=1}^5(x+m_i) \stop
\eeq

\section{Weak coupling monodromies}

As a consistency check of the hyper-elliptic curves describing the
quantum moduli spaces of vacua  which were
constructed in previous sections,
we will compute in this section
the weak coupling monodromies of the curves and compare
to the ones expected on physical grounds.

Let $l$ be a complex line in the moduli space defined by
$s_i=0, i\neq N_c, s_{N_c}=s$.
We consider the massless case, however the generalization
to the massive theories is straightforward.

In the weak coupling limit $s\rightarrow \infty$ ($s \gg \Lambda^{N_c}$)
the hyper-elliptic curve for $N_c,N_f$ theory takes the form
\beq
y^2 =  \(x^{N_c}+ s\)^2
- \Lambda^{2N_c-N_f} x^{N_f} \stop
\label{nnz}
\eeq
The roots of the curve (\ref{nnz}), to first order corrections
in $\Lambda$, are
\beqar
x_{1,l} & = &\varepsilon^l z_1^{1\over N_c} \comma  \nonumber\\
x_{2,l} & = & \varepsilon^l z_2^{1\over N_c}~~~~~l=1,...,N_c \comma
\label{root}
\eeqar
where $\varepsilon = e^{\frac{2 \pi i}{N_c}}$ and
\beqar
z_1 & = & -s\(1 + \Lambda^{\frac{2N_c-N_f}{2}}(-s)^{\frac
{N_f-2N_c}{2N_c}}\) \comma
\nonumber\\
z_2 & = &
-s\(1 - \Lambda^{\frac{2N_c-N_f}{2}}(-s)^{\frac
{N_f-2N_c}{2N_c}}\) \stop
\eeqar

Define the $\alpha_l$ and $\beta_l$ cycles as the
paths
from $x_{1,l}$ to $x_{2,l}$ and
from $x_{1,1}$ to $x_{1,l}$ and back counterclockwise, respectively,
such that the intersection
form of the homology cycles reads $(\beta_k,\alpha_l)=\delta_{kl}$
with $k,l=2,...,N_c$.
Evidently, the $\alpha$ cycles are linearly dependent and
satisfy $\sum_{l=1}^{N_c} \alpha_l =0$.

One way to compute the weak coupling monodromy is to trace the cycles as
$s\rightarrow e^{2 \pi i}s$ around infinity,
as was done in \cite{Alon}.
We will take another route: We will compute the periods of
the curve and study their transformation properties as $s$ traverses
a cycle around infinity.
Denote $A_{kl} = \int_{\alpha_l}\frac{x^{N_c-k}{\rm d}x}{y}$ and
$B_{kl} = \int_{\beta_l}\frac{x^{N_c-k}{\rm d}x}{y}$.
At weak coupling
\footnote{The integrals (\ref{per}) are evaluated by changing variables
to $z=x^{N_c}$ and keeping track of the contours of integration.}
\beqar
B_{kl} = 2\int_{x_{1,1}}^{x_{1,l}}\frac{x^{N_c-k}{\rm d}x}{y} &=&
(\varepsilon^{l(N_c-k+1)}-\varepsilon)
\frac{z_1^{m+\frac{1}{2}}}{
z_2^{\frac{1}{2}}} B\(m+1,\frac{1}{2}\)
F\(\frac{1}{2},m+1,m+\frac{3}{2},
\frac{z_1}{z_2}\), \nonumber\\
A_{kl} = 2\int_{x_{1,l}}^{x_{2,l}}\frac{x^{N_c-k}{\rm d}x}{y} &=&
-\frac{2\pi i}{N_c} \varepsilon^{l(N_c-k+1)} z_1^m F\(-m,\frac{1}{2},1,1-
\frac{z_2}{z_1}\)
\comma
\label{per}
\eeqar
where $B(m,n)$ is the beta function and $m=\frac{1-k}{N_c}$.
Setting $k=l=N_c=2,N_f=0$ in (\ref{per}) yields (\ref{period}).

Let us consider the $k=N_c$ case and expand (\ref{per}) around
infinity in $s$, then recalling (\ref{aap})
\footnote{We have rescaled $a_D$ by factor two.
This change of notation is needed in order
to get the correct monodromies in the sequel.}
\beqar
a^l & \propto & \int A_{N_cl} {\rm d}s  \simeq
2\pi i \varepsilon^l (-s)^{\frac{1}{N_c}}
\comma \nonumber\\
a_D^l & \propto & \int B_{N_cl} {\rm d}s  \simeq
N_c(2N_c-N_f)(\varepsilon-\varepsilon^l) (-s)^{\frac{1}{N_c}}
\log((-s)^{\frac{1}{N_c}}) \stop
\label{aevalu}
\eeqar
First, we can deduce from (\ref{aevalu}) the proportionality
constant in (\ref{aap}).
The classical relation (\ref{symm}) imply that
$a^l=\varepsilon^l (-s)^{\frac{1}{N_c}}$. Thus, comparison to
(\ref{aevalu}) fixes the constant to $\frac{1}{2\pi i}$.

Second, as we shall see now,
(\ref{aevalu}) yields the monodromy at weak coupling as
expected from the one loop  effective action  (\ref{oneloop}).
Denote  $a=(-s)^{\frac{1}{N_c}}$, then (\ref{aevalu}) can be written
as
\beqar
a^l & = &\varepsilon^l a
\comma \nonumber\\
a_D^l & = &
\frac{i}{2\pi}N_c(2N_c-N_f)( \varepsilon^l-\varepsilon) a
\log(a) \comma
\label{oneloop1}
\eeqar
which is consistent with (\ref{F1}) and (\ref{ad}).
Thus we conclude that the periods of
the hyper-elliptic curves that we constructed
have the correct monodromy at infinity, as expected on physical ground.
Note that $a_D$ in (\ref{oneloop1}) can be written as
\beq
a_D^l  =
\frac{i}{2\pi}N_c(2N_c-N_f)( a^l +\sum_{i=2}^{N_c} a^i)
\log(a) \stop
\label{ad1}
\eeq

As $s\rightarrow e^{2 \pi i}s$ on a large cycle in $s$ plane
\beqar
a^l & \rightarrow & a^{l+1}\comma~~~~~l=2,...,N_c-1 \nonumber\\
a^{N_c} & \rightarrow &-
\sum_{i=2}^{N_c}
a^i\comma   \nonumber\\
a_D^l & \rightarrow & a_D^{l+1}-a_D^1 -(2N_c-N_f)(a^{l+1}-a^1)
\comma~~~~~l=2,...,N_c-1 \nonumber\\
a_D^{N_c} & \rightarrow & -a_D^1 +(2N_c-N_f)(a^1+
\sum_{i=2}^{N_c} a^i)\stop
\label{trans}
\eeqar
The matrix representation
of the monodromy at infinity is read from (\ref{trans}). It
takes
the form
\beq
M=\pmatrix{(A^{-1})^t & B \cr 0 & A \cr}_{2g\times 2g} \comma
\label{mon}
\eeq
where
\beq
A=\pmatrix{0 & 1 & 0 & \cdots \cr 0 & 0 & 1 &
\cdots \cr
\vdots & & & 1 \cr
-1 & -1 & \cdots & -1 \cr }_{g\times g}\comma~~~~~
B=-(2N_c-N_f)
\pmatrix{-1 & 1 & 0 & \cdots \cr -1 & 0 & 1 &
\cdots \cr
\vdots & & & 1 \cr
-2 & -1 & \cdots & -1 \cr }_{g\times g}
\stop
\label{AB}
\eeq

In analogy to the $N_c=2$ case \cite{sw2}, the monodromy matrix
(\ref{mon}) can be written as
\beq
M=PT^{2N_c-N_f} \comma
\label{M}
\eeq
where
\beq
P=\pmatrix{(A^{-1})^t & 0 \cr 0 & A \cr}_{2g\times 2g} \comma~~~~~
T=\pmatrix{I & C \cr 0 & I \cr}_{2g\times 2g}
\label{P}
\eeq
and
\beq
C=\pmatrix{2 & 1 & 1 & \cdots \cr 1 & 2 & 1 &
\cdots \cr
\vdots & & 2 & 1 \cr
1 & 1 & \cdots & 2 \cr }_{g\times g} \stop
\eeq
$P$ is the "classical" part of the monodromy associated with Weyl
transformation of the roots of the classical curve (\ref{cl1}).
$T$ is the
"quantum" part of the monodromy associated with the one loop
{\it log} correction to the pre-potential (\ref{F1}).

Finally, let us make a comment on the strong coupling regime.
When $N_f=0$ or $N_f=N_c$, the roots of (\ref{nnz}) can be calculated
exactly. Thus, following the above
calculation, the periods of these curves can be computed exactly.
However, preliminary studies of the strong coupling monodromies
in these cases indicate that there might be a problem with the
computation which is not detected at weak coupling.
\section{The quantum moduli space for $N_f=2N_c$}

\subsection{The general case}

When $N_f=2N_c$ and the bare masses are zero we get conformally
invariant theories.
In these cases the classical relations (\ref{uka}), (\ref{symm})
and (\ref{clf})
are expected to be valid quantum mechanically.
Thus,
\beq
a_D^i =\tau_{ij}a^j \comma
\label{tau}
\eeq
where
\beq
\tau_{ij}=\tau\(\delta_{ij}+1\) \comma
\label{tau1}
\eeq
is the matrix of theta angles and coupling constants
of the theory derived from (\ref{clf}), and
\beq
a_D^j = \tau\(a^j+\sum_{i=2}^{N_c}a^i\)
\stop
\label{adsc}
\eeq
The classical and quantum moduli spaces are identical and are described
by a hyper-elliptic curve with period matrix $\tau_{ij}$ (\ref{tau1})
and
periods as in (\ref{aap}) with  $a^i,a_D^i$ satisfying (\ref{symm})
and (\ref{adsc}).

The structure of the family of curves describing the moduli space
of vacua for $N_f=2N_c$ is encoded in the following:
\newline
{\bf Claim}: The curve for the quantum moduli space for $SU(N_c)$
gauge group with
$N_f=2N_c$ is:
\beq
y^2 = \[{\cal C}_{N_c}(x,l(q)s_k)+P(x,m,q)\]^2-L(q)
\prod_{i=1}^{2N_c}(x+l(q)m_i)  \stop
\label{2nc}
\eeq
$l(q)$ is a modular form satisfying $l(q) \rightarrow 1$ as
$q\rightarrow 0$.
$P(x,m_i,q)$ is modular form satisfying $P \propto q$ as
$q\rightarrow 0$ and a polynomial of degree $N_c$ in
$x$ which is {\it independent} of the moduli $s_k$.
$L(q)$ is a modular form of weight zero satisfying
$L(q) \propto q + O(q^2)$.
\newline
{\bf Proof}: The proof is similar to that of section 5.3.
The main difference is that the dynamically generated scale
$\Lambda$ is replaced by $q$ as defined in (\ref{ql}).
The structure of the first term in (\ref{2nc}) is deduced following the
same argument as in equations (\ref{eqh}) - (\ref{gf}) together with
$R$-symmetry. The factor $g_0$ of (\ref{sol}) gets contributions both
from ${\cal C}_{N_c}(x,l(q)s_k)$ and $P(x,m,q)$.
The modification of all the moduli $s_k$ by the same modular
form $l(q)$ is consistent with $R$-symmetry as well as (\ref{eqh})
and (\ref{sol}).

The structure of the second term follows from the analysis
of the residues (\ref{gres}) of the meromorphic one-form $\lambda$
(\ref{lam1}), which now gets a pre-factor $\frac{1}{l(q)}$.
This implies that $m_i$ must be rescaled by $l(q)$ in order
to ensure that the residues of $\lambda$ be independent of $\tau$.
The behavior of $L(q)$ as $q\rightarrow 0$
is implied by the matching condition (\ref{ql})
when integrating a massive quark.

The compatibility
with (\ref{general}) upon integrating massive
quarks and that of the singularity structure
with the global symmetry suggest that
the curve for $N_f=2N_c$ takes the form
\beq
y^2=\[{\cal C}_{N_c}(x,l(q)s_k)+{L(q)\over4}\sum_{i=0}^{N_c}x^
{N_c-i}t_i(m)\]^2-L(q)\prod_{i=1}^{2N_c}(x+l(q)m_i) \stop
\label{suggest}
\eeq
As for the $N_c < N_f < 2N_c$ cases we do not have a proof of
(\ref{suggest}).
The form of $L(q)$ and
coefficient $l(q)$ of the moduli $s_k$ should be determined such that the
period matrix of the massless curve takes the form (\ref{tau})
and
the periods satisfy (\ref{symm}) and (\ref{tau1}).

Consider now the curve (\ref{suggest})
with the masses being set to zero.
Since equations (\ref{symm}), (\ref{tau}) and (\ref{tau1})
should hold at any generic (non singular) point in the moduli space
let us, for simplicity, use the
complex line $l$ defined previously by
$s_i=0, i\neq N_c,s_{N_c}=s$.

When the masses are set to zero the curve (\ref{suggest}) reads
\beq
y^2=\[\(1+\frac{L}{4}\)x^{N_c} + ls \]^2 -L x^{2N_c} \stop
\label{massless}
\eeq
Its roots take the form
\beqar
x_{1,l} & = &\varepsilon^l z_1^{1\over N_c} \comma  \nonumber\\
x_{2,l} & = & \varepsilon^l z_2^{1\over N_c}~~~~~l=1,...,N_c \comma
\label{root1}
\eeqar
where $\varepsilon = e^{\frac{2 \pi i}{N_c}}$ and
\beq
z_1  =  -\frac{l(q)s}{\(1 + {\sqrt{L}\over2}\)^2} \equiv -\tilde{z_1}
s\comma~~~~~
z_2  =  -\frac{l(q)s}{\(1 - {\sqrt{L}\over2}\)^2} \equiv -\tilde{z_2} s
\stop
\label{z1z2}
\eeq
Evaluating $a^l,a_D^l$ as in section six we get
\beqar
a^l & = & f(q)\varepsilon^l (-s)^{\frac{1}{N_c}}
\comma \nonumber\\
a_D^l & = &
g(q)( \varepsilon^l-\varepsilon)
(-s)^{\frac{1}{N_c}}
\comma
\label{exact}
\eeqar
where  $f$ and $g$ are read  from (\ref{per}).
Note that in contrast to the $N_f<2N_c$ theories, $a_D^l$ do
no get logarithmic corrections.
In order that the classical relation
will hold quantum mechanically we have to require that  $f(q)\equiv 1$,
while in order that (\ref{tau}) be correct we need $\frac{g(q)}{f(q)}=
\tau$.

Let us now suggest possible forms for
$L(q)$ and $l(q)$ which we will verify for the $N_c=2$ case,
but for which we do not have a proof for general $N_c$.
Introduce the theta constant \cite{mumford1}
\beq
\theta \[m_1~~m_2 \] =
\sum_{n\in Z^g} \exp\left\{2 \pi i\[\frac{1}{2}(n+m_1)^t \tau (n+m_1)
+ (n+m_1)^t m_2\]\right\}
\comma
\label{gentheta}
\eeq
where $\tau$ is the period matrix (\ref{tau1}) and
$m_1,m_2$ are dimension $g$ vectors with zeros and halves as
entries.
We suggest that
\footnote{This suggestion is based on the Thomae formula
\cite{mumford2}
for the construction of a hyper-elliptic curve with a prescribed period
matrix, but is not a direct consequence of it.}
\beq
L(q) = \frac{4\theta\[\frac{1}{2}~0\]^4}{\theta [0~0]^4}
\comma~~~~~
l(q)= \frac{\theta\[0~\frac{1}{2}\]^8}{\theta [0~0]^4}
\comma
\label{Ll}
\eeq
where $0$ denotes the zero vector and $\frac{1}{2}$ stands for a vector
with one of its entries being $\frac{1}{2}$ and the others are zeros.
The definition (\ref{gentheta}) with the period matrix (\ref{tau1})
imply that $\theta\[\frac{1}{2}~0\]$ and $\theta\[0~\frac{1}{2}\]$
are independent of which entry is
the $\frac{1}{2}$.
$\theta [0~0]^4$, $\theta\[\frac{1}{2}~0\]^4$ and
$\theta\[0~\frac{1}{2}\]^4$
are modular forms
of $\Gamma_{2,4}$ \footnote{$\Gamma_{2,4} = \left\{
\left( \begin{array}{c}
       A ~~ B \\ C ~~ D
\end{array}  \right) \in Sp(2g,Z)~\biggr\vert~
A, D = 1_{g\times g}~(\mod~2),~
B, C = 0~(\mod~2)\right.$, $4$ divides the diagonals of
$B,C \biggr\rbrace$.} with weight two.

For $N_c=2$ (\ref{Ll}) reduces to
\beq
L(q)= \frac{4\theta_{10}^4}{\theta_{00}^4}
\comma~~~~~
l(q)= \frac{\theta_{01}^8}{\theta_{00}^4}
\comma
\label{Ll2}
\eeq
with
the theta null values \cite{mumford1}
\footnote{In this case the notation amounts to replacing
$\frac{1}{2}$ by $1$. Another notation in the literature is:
$\theta_2(0,q)\equiv \theta_{10}(q)$,
$\theta_3(0,q)\equiv \theta_{00}(q)$
and $\theta_4(0,q)\equiv \theta_{01}(q)$.}
\beqar
\theta_{00}(q) &=& \sum_{n \in Z} q^{n^2} \comma \nonumber\\
\theta_{01}(q) &=& \sum_{n \in Z} (-1)^nq^{n^2} \comma \nonumber\\
\theta_{10}(q) &=& \sum_{n \in Z} q^{\(n+\frac{1}{2}\)^2}
\stop
\label{theta}
\eeqar
$\theta_{00}^4$ , $\theta_{01}^4$ and $\theta_{10}^4$ are modular forms
of $\Gamma_4$ \footnote{$\Gamma_4 = \left\{
\left( \begin{array}{c}
       a ~~ b \\ c ~~ d
\end{array}  \right) \in SL(2,Z)~\biggr|~a,d = 1~(mod~4),~b,c =
0~(mod~4) \right\}$.} with weight two and satisfy the Jacobi identity
\beq
\theta_{00}^4= \theta_{01}^4 +\theta_{10}^4 \stop
\label{idtheta}
\eeq

We verified that for $N_c=2$ the
requirements from the functions $f(q)$ and $g(q)$
are satisfied with $L,l$ of (\ref{Ll2}).
We leave it as an open problem to verify that the requirements
are satisfied for general $N_c$ with $L,l$ of (\ref{Ll}).

\subsection{An example}
{\bf $N_c=2$}: Consider the case $N_f=4$.
The curve $(\ref{2nc})$ takes the form
\beq
y^2 = \[a(q)x^2-b(q)u+c(q) \sum_{i<j}m_im_j
+ d(q) x \sum_{i=1}^4 m_i\]^2 - L(q)
\prod_{i=1}^{4}(x+b(q)m_i)  \stop
\label{24}
\eeq
with the coefficients being modular forms.

Matching to the $N_f=3$ curve when integrating a massive quark
yields
\beqar
a(q)\rightarrow 1 \comma~~~
b(q)\rightarrow 1 \comma~~~
c(q)\rightarrow \frac{q}{8} \comma  \nonumber\\
d(q)\rightarrow \frac{q}{4} \comma~~~
L(q)\rightarrow q \comma
\label{limits}
\eeqar
as $q \rightarrow 0$.

The global symmetries together with (\ref{limits}) yield
\beq
a(q)=1+\frac{L}{4} \comma~~~
c(q)=\frac{L}{8}\comma~~~
d(q)=\frac{L}{4} \stop
\label{values}
\eeq
$L(q)$ and  $b(q)$ are determined by requiring that in the massless
case the periods of the curve are such that the classical relation
$a=\sqrt{u}$ and (\ref{tau}) are satisfied.
These yield $L(q)$ as in (\ref{Ll2}) and
\beq
b(q)= \frac{\theta_{01}^8}{\theta_{00}^4}=l(q) \stop
\label{b}
\eeq
Thus, the $N_c=2,N_f=4$ curve reads
\beq
y^2 = \[\(1+\frac{L(q)}{4}\)x^2- l(q)u+
\frac{L(q)}{8} \sum_{i<j}m_im_j
+ \frac{L(q)}{4} x \sum_{i=1}^4 m_i\]^2 - L(q)
\prod_{i=1}^{4}(x+l(q)m_i)  \comma
\label{24f}
\eeq
with $L(q)$ and $l(q)$ given by (\ref{Ll2}).

\section{Discussion and conclusions}

In this paper we
constructed the hyper-elliptic curves which describe the quantum
moduli spaces of vacua of $N=2$ supersymmetric $SU(N_c)$ gauge theories with
$N_f$ flavors of quarks in the fundamental representation.
We showed that
the curves for $N_f<N_c$ are completely determined by $R$-symmetry, the form
of instanton corrections and the requirement
for the correct classical singularity structure.
The compatibility of the residue calculations with the
BPS formula as well as the correct weak coupling monodromy
provide further support to the results.

As expected, the complete specification of the
curves for $N_f\ge N_c$ is more complicated.
As in the $SU(2)$ case \cite{sw2}, the residues of the meromorphic
one-form $\lambda$ provide strong constraints on the structure
of the curves. Together with the other principles, discussed in
section 2.4, we worked out the structure of the curves, up to certain
unknown constant coefficients.

We have not fully exploited the relation between the
global symmetries and the singularity structure.
However, an observation that we made in section 2.3 on the form of
the discriminants suggested a complete
determination of all the unknown coefficients.
It will be interesting to establish this observation on a firm basis,
and to fully understand the physics underlying it.

Weak coupling monodromies were computed for all the
curves and were shown to coincide with what is expected on physical
grounds, thus providing a check on the results.
We left the calculation of the strong coupling monodromies
for the future. This will clearly be needed in order
to extract the physics of these theories.
Along the way we derived the
exact metrics on the quantum moduli spaces as well as the exact
spectrum of stable massive states.

We found an example of a novel symmetry of a quantum moduli
space: Invariance under the exchange of a moduli parameter
and the bare mass.
This implies a sort of duality that relates theories with the same
gauge group and different vacua, to be contrasted with the duality
of \cite{Seiberg} that relates theories with different gauge groups.
Physically such a symmetry is surprising since it relates
bare parameters of the classical Lagrangian combined with a dynamically
generated mass scale to the vacuum expectation value of the scalar fields.
We expect more symmetries of this type to appear in these
theories and it will be important to find their general form.

An open question is to find the singular points that correspond
to $N=1$ vacua. As discussed in \cite{DS} for
the $N_f=0$ theories, these points are
associated with curves of degree $2N_c$ with $N_c-1$ pairs of identical
branch points.
A preliminary check of the massless
$N_c=3$ curves with $0 < N_f \leq 6$, that have
been constructed in this paper, did not reveal such points.
Evidently, we expect to see these points in the massive cases.
Identifying these points will clearly shed more light on the physics
of these theories in the strong coupling regime.
Another direction is to study the theories at the singular points that
correspond to non mutually local massless fields \cite{DA}.

In the appendix we applied our method for constructing the quantum
moduli spaces, to $N=1$ supersymmetric $SU(N_c)$ gauge theories
with a single matter field in the adjoint representation
of the gauge group and $N_f$ matter fields in the fundamental
representation. This generalizes the constructions of the quantum
moduli spaces of $N=2$ to general mass matrix and Yukawa couplings.
It is clear that the method that we use for constructing the curves
is rather general and can be applied in a straightforward manner
to a vast number of $N=1$ theories in the coulomb phase
with different matter content.
Another, less straightforward,
direction for generalization is to include other gauge
invariant moduli such as meson operators and to construct the quantum moduli
spaces for the theories discussed in \cite{Kutasov,ASY,SK}.

The ability to extract exact results in the theories that were studied
in this paper points to an underlying integrable
structure \cite{Mor}. In particular, one expects that the pre-potential
will be
related to
a $\tau$ function of some integrable hierarchy and that the variety
describing the quantum moduli space of vacua
will arise as a solution to a non-linear
integrable equation of the hierarchy. Revealing these structures
may provide us with powerful computational tools
for these four-dimensional models.

\section*{Acknowledgements}
We would like to thank
A.~Morozov, A.~Schwimmer, N.~Seiberg and
V.~Vinnikov for helpful discussions.

\appendix{On $N=1$ quantum moduli spaces of vacua}

The procedure applied in this paper
for constructing the curves, which describe the quantum moduli spaces
of $N=2$ supersymmetric gauge theories is rather general.
It can be used, for instance, in order to construct the
curves which describe the effective Abelian gauge field couplings in the
Coulomb
phase
of various $N=1$ theories with different matter content.

As an example, we will construct in this appendix
the hyper-elliptic curves which describe the
quantum moduli spaces of $N=1$ supersymmetric $SU(N_c)$ gauge theories
with one flavor in the adjoint representation of the
gauge group and $N_f$
flavors in the fundamental representation,
and with general mass matrix $m_{ij}$
and Yukawa couplings $\lambda_{ij}$
\footnote{Setting $m_{ij}=diag[m_1,...,m_{N_f}]$ and
$\lambda_{ij}=\delta_{ij}$ for $N_c \neq 2$ yields the
$N=2$ theories.}.
In these cases the
super-potential takes the form
\beq
W=\lambda_{ij} \tilde Q^i \Phi Q^j + m_{ij}\tilde Q^i Q^j  \comma
\label{n1superpot}
\eeq
with $i$ being the flavor index and color indices suppressed.
Consider first the $N_f<N_c$ theories.
As we have seen in section three, $R$-symmetry, instanton corrections
and the classical singularity structure determine the quantum
moduli space curve completely.
The super-potential (\ref{n1superpot}) implies
that there is classically a massless state whenever
\beq
\Delta_{N_f,N_c}=\det_{\alpha\beta}\det_{ij}(\lambda_{ij}\phi_{\alpha\beta}
+m_{ij}\delta_{\alpha\beta})   \comma
\label{n1cladis}
\eeq
vanishes.
Thus, the classical discriminant should have the form (\ref{n1cladis}).
The function $G(x,m_i)$ of (\ref{G})
generalizes and for $N_c>2$ takes the form of
the characteristic polynomial of the matrix $\lambda^{-1}M$
\beq
G(x,m,\lambda)=\det(\lambda x+m)=
\det\lambda\det(x+\lambda^{-1}m)=
\det\lambda\sum_{i=1}^{N_f}t_ix^{N_f-i}  \stop
\eeq
where $t_k$ are the symmetric functions of the matrix $\lambda^{-1}m$
defined
by the Newton formula (\ref{Newton})
with $u_k$ being $\Tr(\lambda^{-1}m)^k$.
In analogy to (\ref{gm}), the curves describing the
quantum moduli spaces for $N_f<N_c$ where $N_c>2$ take the form
\beq
y^2=\det_{\alpha\beta}(x-\phi)^2-\Lambda^{2N_c-N_f}
\det_{ij}\lambda\det_{ij}(x+\lambda^{-1}M)
\stop
\eeq

The $N_c=2$ case is more subtle, as a consequence of the fact
that the fundamental representation is pseudo-real.
The super-potential in this case reads
\beq
W=\lambda_{ij} \tilde Q^i \Phi Q^j + m_{ij}\tilde Q^i Q^j  \comma
\label{n1superpot1}
\eeq
where $\lambda$ and $m$ are $2N_f\times2N_f$ symmetric and antisymmetric
matrices, respectively.
The classical discriminant is given by
\beq
\Delta_{N_f,2}=\det\lambda\sum_{i=1}^{N_f}t_{2i}(\lambda^{-1}M)(-u)^{N_f-i}
\comma
\eeq
and the curve describing the quantum moduli space of $N_c=2, N_f=1$
is
\beq
y^2=(x^2-u)^2-\Lambda^3 {\rm Pf}(\lambda)(x+ {\rm Pf}(\lambda^{-1}m))
\stop
\eeq
The higher flavor $N_c=2$ curves can be constructed in a complete
analogy to the $N=2$ models.
For instance,
the curve for $N_c=2, N_f=2$ is given by
\beq
y^2=\(x^2-u+{\Lambda^2{\rm Pf}\lambda\over8}\)^2
-\Lambda^2{\rm Pf}\lambda\(x^2+x\sqrt{2{\rm Pf}(\lambda^{-1}m)-t_2}
+{\rm Pf}(\lambda^{-1}m)\).
\eeq

Following (\ref{general}), we suggest that
the curves for $N_f\geq N_c$ with $N_c>2$ take the form
\beq
y^2=\[\det_{\alpha\beta}(x-\phi)
+{\Lambda^{2N_c-N_f}\det_{ij}\lambda\over4}\sum_{i=0}^{N_f-N_c}t_i
x^{N_f-N_c-i}\]^2-\Lambda^{2N_c-N_f}\det_{ij}\lambda\det_{ij}(x+\lambda^{-1}M).
\eeq

The results presented in this appendix generalize those of \cite{EFGR}
to $N_c>2$.

\end{document}